\documentstyle[a4,epsfig,amsfonts]{article}

\input amssym.def
\input amssym
\title{Free massive fermions inside the quantum discrete sine-Gordon
model}
\author{Nadja Kutz\addtocounter{footnote}{-1}\thanks{Supported 
by the Deutsche Forschungsgemeinschaft, Sonderforschungsbereich 288}%
\thanks{e-mail: nadja@math.tu-berlin.de}
\\ Technische Universit\"at Berlin\\
Sekr. MA 8-5, Str. des 17. Juni 136 \\
10623 Berlin, Germany}
\date{24.10.96}

\newcommand{\dgleich}{\raisebox{0.07ex}{:}\!=}
\newcommand{\real}{{\Bbb R}}
\newcommand{\REELL}{{\Bbb R}}

\newcommand{\NATUR}{{\Bbb N}}

\newcommand{\KOMPLEX}{\Bbb C}

\newcommand{\GANZ}{\Bbb Z}
\newcommand{\comment}[1]{}
\newtheorem{Theorem}{Theorem}[section]
\newtheorem{Proposition}[Theorem]{Proposition}
\newtheorem{Lemma}[Theorem]{Lemma}
\newtheorem{Demand}[Theorem]{Demand}
\newtheorem{Corollary}[Theorem]{Corollary}

\newcommand{\boxlisttype}{\bf}
\newcommand{\boxlistlabel}[1]{\boxlisttype #1\hfil}
\newenvironment{boxlist}[2]{%
\renewcommand{\boxlisttype}{#1}%
\begin{list}{#2}{%
\settowidth{\labelwidth}{\boxlistlabel{#2}}%
\setlength{\labelsep}{.5em}%
\setlength{\leftmargin}{\labelwidth}%
\addtolength{\leftmargin}{\labelsep}%
}}{%
\end{list}}
\newcommand{\bbl}[1]{\begin{boxlist}{\rm}{#1}}
\newcommand{\bhin}{\begin{boxlist}{\it}{Hinweis:}\item[Hinweis:]}
\newcommand{\bbem}{\begin{boxlist}{\it}{Bemerkung:}\item[Bemerkung:]}
\newcommand{\ebl}{\end{boxlist}}
\newcommand{\beqn}{\begin{equation} }
\newcommand{\eeqn}{\end{equation}}

\newcommand{\beqna}{\begin{eqnarray}}
\newcommand{\beqnao}{\begin{eqnarray*}}
\newcommand{\eeqna}{\end{eqnarray}}
\newcommand{\eeqnao}{\end{eqnarray*}}
\newcommand{\ba}{\begin{array}}
\newcommand{\ea}{\end{array}}
\newcommand{\unity}{{\setlength{\unitlength}{1em}
                     \begin{picture}(0.75,1)
                     \put(0,0){$1$}
                     \put(0.34,0){\line(0,1){0.65}}
                     \end{picture}
                   }}
\begin{document}

\begin{titlepage}
\maketitle

\begin{abstract}
We extend the notion of space shifts introduced in \cite{FV3} for 
certain quantum light cone lattice equations of sine-Gordon type at
root of unity (e.g. \cite{FV1}\cite{FV2}\cite{BKP}\cite{BBR}).
As a result we obtain a compatibility equation for the roots
of central elements within the algebra of observables (also
called current algebra). The equation which is obtained by  exponentiating
these roots is exactly the evolution equation for the "classical background"
as described in \cite{BBR}.

As an application for the introduced constructions, a one to one 
correspondence between a special case of the quantum light cone
lattice equations of sine-Gordon type and free massive fermions 
on a lattice as constructed in \cite{DV} is derived.
\end{abstract}

\end{titlepage}

\section{Introduction}
Let us consider a class of integrable lattice systems 
defined by an evolution equation of the following type:
\beqn \label{eq:evol}
 g_u = V'(g_l-g_r) + g_d=0,
\eeqn
\noindent where $u$, $d$, $l$ and $r$ denote up, down, left and
right, respectively and $V'(x)$ is the derivative of $V(x): \real
\rightarrow \real$. If we start with quasiperiodic 
initial data $g_i$ along a Cauchy path ${\cal C}$ on a light cone  
lattice the local
evolution given by (\ref{eq:evol}) will determine the function $g$
on the whole lattice.

\begin{center}
\begin{picture}(0,0)%
\epsfig{file=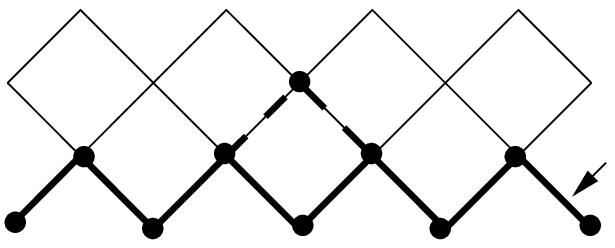}%
\end{picture}%
\setlength{\unitlength}{0.00083300in}%
\begingroup\makeatletter\ifx\SetFigFont\undefined
\def\x#1#2#3#4#5#6#7\relax{\def\x{#1#2#3#4#5#6}}%
\expandafter\x\fmtname xxxxxx\relax \def\y{splain}%
\ifx\x\y   
\gdef\SetFigFont#1#2#3{%
  \ifnum #1<17\tiny\else \ifnum #1<20\small\else
  \ifnum #1<24\normalsize\else \ifnum #1<29\large\else
  \ifnum #1<34\Large\else \ifnum #1<41\LARGE\else
     \huge\fi\fi\fi\fi\fi\fi
  \csname #3\endcsname}%
\else
\gdef\SetFigFont#1#2#3{\begingroup
  \count@#1\relax \ifnum 25<\count@\count@25\fi
  \def\x{\endgroup\@setsize\SetFigFont{#2pt}}%
  \expandafter\x
    \csname \romannumeral\the\count@ pt\expandafter\endcsname
    \csname @\romannumeral\the\count@ pt\endcsname
  \csname #3\endcsname}%
\fi
\fi\endgroup
\begin{picture}(2980,1271)(1221,-4036)
\put(2026,-3511){\makebox(0,0)[lb]{\smash{\SetFigFont{10}{12.0}{rm}$g_l$}}}
\put(4201,-3511){\makebox(0,0)[lb]{\smash{\SetFigFont{10}{12.0}{rm}Cauchypath ${\cal C}$}}}
\put(3122,-3477){\makebox(0,0)[lb]{\smash{\SetFigFont{10}{12.0}{rm}$g_r$}}}
\put(2598,-3011){\makebox(0,0)[lb]{\smash{\SetFigFont{10}{12.0}{rm}$g_u$}}}
\put(2598,-4003){\makebox(0,0)[lb]{\smash{\SetFigFont{10}{12.0}{rm}$g_d$}}}
\end{picture}
\end{center}

\noindent Integrable lattice systems of the above type
 had been thoroughly discussed e.g. in \cite{BRST,NCP,S,V,FV2,CN,EK})

\noindent In \cite{EK} it was shown that it is possible to derive the above 
equation as an equation of motion from an explicitly given action.
Moreover - using
covariant phase space techniques  -  it is possible
to derive the symplectic structure belonging to the above model 
via variation of this action. As a consequence one obtains unique
Poisson commutation rules for the above variables which were 
first stated in \cite{FV2}. The so obtained Poisson commutation rules
are not local in space, while the induced commutation rules for
the difference variables $g_l-g_r$ (also called current variables)
are. Therefore 
the difference variables were looked at as referring to physical 
observables and consequently a lot of attention was drawn to them.
(see e.g. \cite{BKP,FV1,BBR,CN}).

Nevertheless it may be worthwile, also
for physical reasons,  to study the above variables despite their
unpleasant nonultralocal commutation relations. 
An important step in this direction was made in the work of 
Faddeev and Volkov \cite{FV3}, were quantized models 
of the above variables were studied . 
Moreover in this work shifts in space direction as automorphisms on 
the quantized variables were constructed under the assumption of 
special monodromies.

We will extend their definition in the root of unity case
to more general monodromies and 
derive as a consequence of this extension in connection with the
corresponding quantum evolution a compatibility equation 
for the roots of central elements within the algebra of the 
current variables. The equation which is obtained by  exponentiating
these roots is exactly the evolution equation for the "classical background"
as described in \cite{BBR}.

As a concrete example for the above construction of space and time
shifts we will finally restrict ourselves to a subalgebra
of the algebra generated by the quantized quasiperiodic variables $g$.
The generators of this subalgebra satisfy canonical anticommutation 
rules (CAR). The quantum analog of the evolution  in (\ref{eq:evol})
reduced to this subalgebra for the case of $V(x)$ being the potential
of the sine-Gordon model will be isomorphic to the evolution of
free massive fermions on the lattice viewed as a special case
of the massive Thirring model  constructed by Destri and de Vega
\cite{DV} (see also \cite{TS} and others cited there).

As a result we will may be understand more about the true nature of the 
famous sine-Gordon - massive Thirring model equivalence
(see e.g. \cite{C,KM}).

\section{The phase space}
\subsection{Vertex variables}
\noindent 
A (spatially) periodic light cone lattice $L_{2p}$ (see also fig. 1) with 
period $2p$ may be viewed
as $ L/\GANZ$, where $\GANZ$ acts on the infinite light cone 
lattice $L$ by shifts by $2p$ in 
space-like direction (cf. fig. 1). A quasi-periodic field 
is a mapping 
\[ g: L \rightarrow \REELL\]
with 
\[ g_{t,i+2p} - g_{t,i} = g_{t,i+2p+2}- g_{t,i+2}  ~~ \forall i \in \GANZ,\]
i.e., there are two (space independent) monodromies $m^{(1)}_t, m^{(2)}_t$
defined by 
\beqn \label{def:Vmon}
m_{t}^{(i)} = g_{t,2p+ 2k +1-i} - g_{t,2k+ 1-i} 
\eeqn
 for an arbitrary $k \in
\GANZ$.

\begin{center}  
\begin{picture}(0,0)%
\epsfig{file=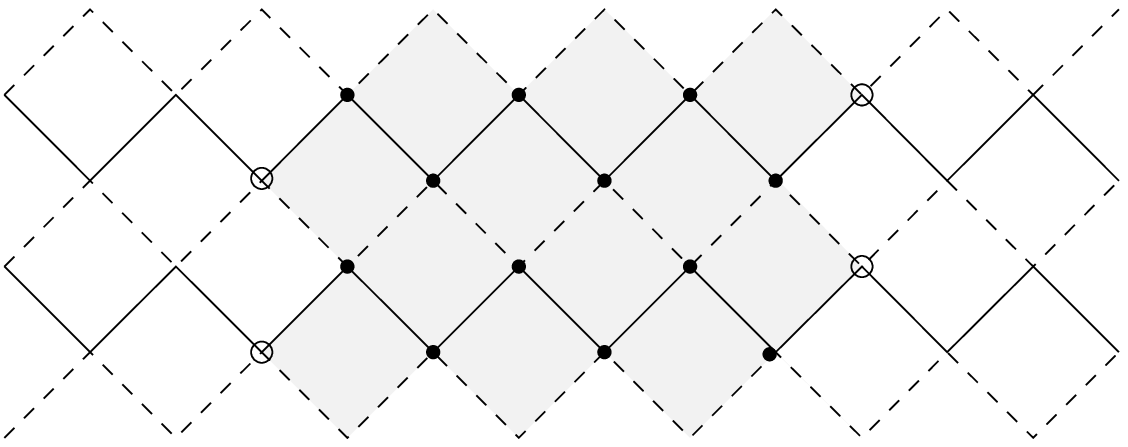}%
\end{picture}%
\setlength{\unitlength}{0.00083300in}%
\begingroup\makeatletter\ifx\SetFigFont\undefined
\def\x#1#2#3#4#5#6#7\relax{\def\x{#1#2#3#4#5#6}}%
\expandafter\x\fmtname xxxxxx\relax \def\y{splain}%
\ifx\x\y   
\gdef\SetFigFont#1#2#3{%
  \ifnum #1<17\tiny\else \ifnum #1<20\small\else
  \ifnum #1<24\normalsize\else \ifnum #1<29\large\else
  \ifnum #1<34\Large\else \ifnum #1<41\LARGE\else
     \huge\fi\fi\fi\fi\fi\fi
  \csname #3\endcsname}%
\else
\gdef\SetFigFont#1#2#3{\begingroup
  \count@#1\relax \ifnum 25<\count@\count@25\fi
  \def\x{\endgroup\@setsize\SetFigFont{#2pt}}%
  \expandafter\x
    \csname \romannumeral\the\count@ pt\expandafter\endcsname
    \csname @\romannumeral\the\count@ pt\endcsname
  \csname #3\endcsname}%
\fi
\fi\endgroup
\begin{picture}(5373,2583)(1427,-4747)
\put(3547,-3051){\makebox(0,0)[lb]{\smash{\SetFigFont{7}{8.4}{rm}$g_{2t+1,1}$}}}
\put(5194,-3051){\makebox(0,0)[lb]{\smash{\SetFigFont{7}{8.4}{rm}$g_{2t+1,2p-1}$}}}
\put(2313,-3460){\makebox(0,0)[lb]{\smash{\SetFigFont{7}{8.4}{rm}$g_{2t,-2}$}}}
\put(6015,-3051){\makebox(0,0)[lb]{\smash{\SetFigFont{7}{8.4}{rm}$g_{2t+1,2p+1}$}}}
\put(3135,-3460){\makebox(0,0)[lb]{\smash{\SetFigFont{7}{8.4}{rm}$g_{2t,0}$}}}
\put(2724,-3051){\makebox(0,0)[lb]{\smash{\SetFigFont{7}{8.4}{rm}$g_{2t+1,-1}$}}}
\put(2313,-2638){\makebox(0,0)[lb]{\smash{\SetFigFont{7}{8.4}{rm}$g_{2t+2,-2}$}}}
\put(3135,-2638){\makebox(0,0)[lb]{\smash{\SetFigFont{7}{8.4}{rm}$g_{2t+2,0}$}}}
\put(3958,-2638){\makebox(0,0)[lb]{\smash{\SetFigFont{7}{8.4}{rm}$g_{2t+2,2}$}}}
\put(5606,-2638){\makebox(0,0)[lb]{\smash{\SetFigFont{7}{8.4}{rm}$g_{2t+2,2p}$}}}
\put(3958,-3460){\makebox(0,0)[lb]{\smash{\SetFigFont{7}{8.4}{rm}$g_{2t,2}$}}}
\put(3346,-3676){\makebox(0,0)[lb]{\smash{\SetFigFont{7}{8.4}{rm}$x_{2t,1}$}}}
\put(3721,-3661){\makebox(0,0)[lb]{\smash{\SetFigFont{7}{8.4}{rm}$x_{2t,2}$}}}
\put(6729,-3270){\makebox(0,0)[lb]{\smash{\SetFigFont{9}{10.8}{rm}${\cal C}_{2t+1}$}}}
\put(6750,-3605){\makebox(0,0)[lb]{\smash{\SetFigFont{9}{10.8}{rm}${\cal C}_{2t}$}}}
\put(6015,-3872){\makebox(0,0)[lb]{\smash{\SetFigFont{7}{8.4}{rm}$g_{2t-1,2p+1}$}}}
\put(2881,-3676){\makebox(0,0)[lb]{\smash{\SetFigFont{7}{8.4}{rm}$x_{2t,0}$}}}
\put(5606,-3460){\makebox(0,0)[lb]{\smash{\SetFigFont{7}{8.4}{rm}$g_{2t,2p}$}}}
\put(5194,-3872){\makebox(0,0)[lb]{\smash{\SetFigFont{7}{8.4}{rm}$g_{2t-1,2p-1}$}}}
\put(3547,-3872){\makebox(0,0)[lb]{\smash{\SetFigFont{7}{8.4}{rm}$g_{2t-1,1}$}}}
\put(2674,-3872){\makebox(0,0)[lb]{\smash{\SetFigFont{7}{8.4}{rm}$g_{2t-1,-1}$}}}
\put(4051,-3286){\makebox(0,0)[lb]{\smash{\SetFigFont{7}{8.4}{rm}$x_{2t+1,3}$}}}
\put(3286,-3256){\makebox(0,0)[lb]{\smash{\SetFigFont{7}{8.4}{rm}$x_{2t+1,1}$}}}
\put(3976,-4711){\makebox(0,0)[lb]{\smash{\SetFigFont{12}{14.4}{rm}Figure 1}}}
\end{picture}
\end{center}

\noindent 
We define an evolution of the following form, where the indices $t,k$
will be adapted to Figure 1, i.e. $k-t$ is chosen to be even:
\beqn
g_{t+1,k+1} = V'(g_{t,k}-g_{t,k+2}) + g_{t-1,k+1}
\qquad m_{t+1}^{(i)} = const \quad m_{t}^{(i)} = const
\label{eq:evgt}\eeqn
For a set of initial values  $ {\cal I}_{2t}^g=
\bigl\{(g_{2t-1,k})_{k \in \{-1,\ldots 2p-1\}},
(g_{2t,k})_{k \in \{0,\ldots 2p\}}\bigr\}$ or 
$ {\cal I}_{2t+1}^g =
\bigl\{(g_{2t,k})_{k \in \{0,\ldots 2p\}},
(g_{2t+1,k})_{k \in \{-1,\ldots 2p-1\}}\bigr\}$, ($t \in \GANZ$ fixed)
 along an elementary Cauchy zig zag (see fig. 1) the above evolution
equations define quasiperiodic field values at all other times. 
A quasiperiodic field $g$
obtained in such a way will be called a solution to the above 
evolution (\ref{eq:evgt}). Hence the quasiperiodic initial values  
$ {\cal I}_t$ can  be interpreted as a (global) coordinate chart 
$g_t: {\cal P} \rightarrow \REELL^{2p+2}$ on 
the set of all solutions 
${\cal P}=\{g|_{solution}\}$ via the identification:
$$
g_{t,i}(g) = g_{t,i} \in \REELL.
$$
The set of all solutions to a given evolution shall be called 
covariant phase space ${\cal P}$. It is now possible to define a 
translational invariant action
on the set of quasiperiodic fields whose variation gives the 
above field equations as well as a time independent, translational invariant 
 symplectic structure
\cite{EK} which in the turn defines the following 
Poisson  relations:
\begin{equation}
\begin{array}{ccll}
 \{g_{t,i}, g_{t,k}\} &=& 0 , &\mbox{ if $i-k$ even}\\[1mm]
 \{g_{t,i}, g_{\tilde{t},k}\} &=& 1 ,  &\mbox{ if $i-k$ odd, $i<k,\; |i-k|<2p$},\\[1mm]
 \{g_{t,i}, m^{(k)}_t\}& =& 0 &\mbox{ ($i-k$ odd),} \\[1mm]
 \{g_{t,i}, m^{(k)}_{\tilde{t}}\} &=& 2 &\mbox{  ($i-k$ even),}\\
 \{m^{(1)}_t,m^{(2)}_{\tilde{t}} \} & = & 0
\end{array}
\label{eq:poisVertex}\end{equation}
where $|t-\tilde{t}| = 1.$

\noindent Since the quasiperiodic variables $ \bigl\{g_{t,k}\bigr\}$ 
are associated to the vertices of the Minkowski lattice  $L$ they
will be shortly called {\em vertex} variables. Since the monodromies as dynamical
variables are by virtue of the evolution equations time independent we will skip from now on
their time index $t$ and keep in mind that 
$m^{(1)} = m^{(1)}_{2t}$ and  $m^{(2)} = m^{(2)}_{2t-1}$.

\subsection{Edge variables}
\noindent Define the following variables for ($k-t$) even:
\beqna
x_{t,k} &\dgleich& g_{t,k} + g_{t-1,k-1} \nonumber\\
x_{t,k-1}&\dgleich& g_{t-1,k-1} + g_{t,k-2}
\label{def:x}
\eeqna
\begin{center}
\begin{picture}(0,0)%
\epsfig{file=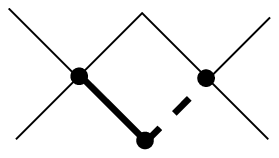}%
\end{picture}%
\setlength{\unitlength}{0.00083300in}%
\begingroup\makeatletter\ifx\SetFigFont\undefined
\def\x#1#2#3#4#5#6#7\relax{\def\x{#1#2#3#4#5#6}}%
\expandafter\x\fmtname xxxxxx\relax \def\y{splain}%
\ifx\x\y   
\gdef\SetFigFont#1#2#3{%
  \ifnum #1<17\tiny\else \ifnum #1<20\small\else
  \ifnum #1<24\normalsize\else \ifnum #1<29\large\else
  \ifnum #1<34\Large\else \ifnum #1<41\LARGE\else
     \huge\fi\fi\fi\fi\fi\fi
  \csname #3\endcsname}%
\else
\gdef\SetFigFont#1#2#3{\begingroup
  \count@#1\relax \ifnum 25<\count@\count@25\fi
  \def\x{\endgroup\@setsize\SetFigFont{#2pt}}%
  \expandafter\x
    \csname \romannumeral\the\count@ pt\expandafter\endcsname
    \csname @\romannumeral\the\count@ pt\endcsname
  \csname #3\endcsname}%
\fi
\fi\endgroup
\begin{picture}(1433,888)(-299,-3097)
\put(278,-3064){\makebox(0,0)[lb]{\smash{\SetFigFont{9}{10.8}{rm}$g_{t-1,k-1}$}}}
\put( 28,-2824){\makebox(0,0)[lb]{\smash{\SetFigFont{9}{10.8}{rm}$x_{t,k-1}$}}}
\put(701,-2842){\makebox(0,0)[lb]{\smash{\SetFigFont{9}{10.8}{rm}$x_{t,k}$}}}
\put(946,-2614){\makebox(0,0)[lb]{\smash{\SetFigFont{9}{10.8}{rm}$g_{t,k}$}}}
\put(-299,-2611){\makebox(0,0)[lb]{\smash{\SetFigFont{9}{10.8}{rm}$g_{t,k-2}$}}}
\put(676,-2386){\makebox(0,0)[lb]{\smash{\SetFigFont{9}{10.8}{rm}$x_{t+1,k}$}}}
\put( 76,-2386){\makebox(0,0)[lb]{\smash{\SetFigFont{9}{10.8}{rm}$x_{t+1,k-1}$}}}
\end{picture}
\end{center}

\noindent The  variables $ \bigl\{x_{t,k}\bigr\}$ 
are associated to the edges of the Minkowski lattice $L$. 
They will be shortly called {\em edge} variables.

\noindent
A set of initial vertex values $ {\cal I}_{2t+1}^g$ or 
$ {\cal I}_{2t}^g$, respectively defines the initial edge values
$ {\cal I}_{2t+1}^x =
\bigl\{(x_{2t+1,k})_{k \in \{0,\ldots 2p\}}\bigr\}$ or
$ {\cal I}_{2t}^x = (x_{2t,k})_{k \in \{0,\ldots 2p\}}\bigr\}$. 

\noindent  The edge variables are still quasiperiodic. Their monodromy
is the sum of the two monodromies of the vertex variables :
$$
x_{t,k+2p} = x_{t,k} + \underbrace{m^{(1)} + m^{(2)}}_{\dgleich m^x}
$$
 The induced Poisson commutation rules are 
\beqna
\{x_{t,k}, x_{t,k+n}\} &=& 2 \qquad n \in \{1 \ldots 2p-1\}\quad mod\, 2p\\
\{x_{t,k}, m_t^x\} &=& 4
\eeqna

\noindent One obtains evolution equations in the 
edge variables by virtue of definition (\ref{def:x}) and
the vertex evolution equations (\ref{eq:evgt}), they read as:
\beqna
x_{t+1,k} &=& V'(x_{t,k-1} - x_{t,k}) + x_{t,k}\\
x_{t+1,k-1} &=& V'(x_{t,k-1} - x_{t,k}) + x_{t,k-1}\\
m^x &=& const.
\eeqna

\noindent We will call these equations edge evolution equations.

\subsection{Face variables}
Define the following fields ($k-t$ even):
$$p_{t,k-1} \dgleich g_{t,k-2} - g_{t,k}$$

\begin{center}
\begin{picture}(0,0)%
\epsfig{file=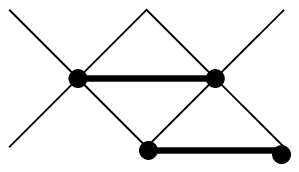}%
\end{picture}%
\setlength{\unitlength}{0.00083300in}%
\begingroup\makeatletter\ifx\SetFigFont\undefined
\def\x#1#2#3#4#5#6#7\relax{\def\x{#1#2#3#4#5#6}}%
\expandafter\x\fmtname xxxxxx\relax \def\y{splain}%
\ifx\x\y   
\gdef\SetFigFont#1#2#3{%
  \ifnum #1<17\tiny\else \ifnum #1<20\small\else
  \ifnum #1<24\normalsize\else \ifnum #1<29\large\else
  \ifnum #1<34\Large\else \ifnum #1<41\LARGE\else
     \huge\fi\fi\fi\fi\fi\fi
  \csname #3\endcsname}%
\else
\gdef\SetFigFont#1#2#3{\begingroup
  \count@#1\relax \ifnum 25<\count@\count@25\fi
  \def\x{\endgroup\@setsize\SetFigFont{#2pt}}%
  \expandafter\x
    \csname \romannumeral\the\count@ pt\expandafter\endcsname
    \csname @\romannumeral\the\count@ pt\endcsname
  \csname #3\endcsname}%
\fi
\fi\endgroup
\begin{picture}(1618,945)(1021,-3469)
\put(1758,-3014){\makebox(0,0)[lb]{\smash{\SetFigFont{10}{12.0}{rm}$p_{t,k-1}$}}}
\put(1021,-2907){\makebox(0,0)[lb]{\smash{\SetFigFont{10}{12.0}{rm}$g_{t,k-2}$}}}
\put(2385,-2892){\makebox(0,0)[lb]{\smash{\SetFigFont{10}{12.0}{rm}$g_{t,k}$}}}
\put(2176,-3436){\makebox(0,0)[lb]{\smash{\SetFigFont{10}{12.0}{rm}$p_{t-1,k}$}}}
\end{picture}
\end{center}

\noindent 
The difference variables $ \bigl\{p_{t,k}\bigr\}$ 
 are associated to the faces
of the Minkowski lattice $L$. They will be shortly called {\em face}
variables. 
A set of initial vertex values $ {\cal I}_{2t+1}^g$ or 
$ {\cal I}_{2t}^g$, respectively defines the initial face values
$ {\cal I}_{2t}^p =
\{(p_{2t-1,2k})_{k \in \{0..p-1\}}$, $(p_{2t,2k+1})_{k \in \{0..p-1\}}\}$
or 
$ {\cal I}_{2t+1}^p =
\{(p_{2t,2k+1})_{k \in \{0..p-1\}}$, $(p_{2t+1,2k})_{k \in \{0..p-1\}}\}.$

\noindent Note that in particular $p_{t,k-1} = x_{t,k-1} - x_{t,k} 
= x_{t+1,k-1} - x_{t+1,k}$ ($k-t$ even).
The above variables are periodic since the monodromies cancel out:
$$
p_{t,2p+k-1} = p_{t,k-1} = g_{t,k-2} + m^{(i)} - (g_{t,k} + m^{(i)}).
$$

\noindent The induced Poisson relations between these variables are
ultralocal in space:
\beqna
\{p_{t,j},p_{\tilde{t},j+1}\} &=& 2\\
\{p_{t,j},p_{t,j+1+k}\} &=& 0 \quad \mbox{for}\quad k \in \{1 \ldots 2p-3\} 
\eeqna
 where $|t-\tilde{t}| = 1 $.

\noindent 
The evolution equations in terms of the face variables read as
$$
p_{t+1, k} = V'(p_{t,k-1}) - V'(p_{t,k+1}) + p_{t-1,k}.
$$
We will call these equations face evolution equations.

\noindent The introduction of the above variables refers
 to a reduction of phase space as explained
in \cite{EK}.

\subsection{Relations to the sine-Gordon model}

Let the halfperiod p of the lattice be even, i.e.
$p = 2s$. In \cite{EK} it was shown that the action, which describes the
above dynamical systems is invariant under a redefinition 
of $g \leadsto -g$ along every second diagonal of the Minkowski lattice,
i.e. all the above structure is preserved under such a redefinition.

\noindent Choosing
\beqn
V'(x) = -i \, ln (\frac{1+ k e^{ix}}{k + e^{ix}})
\label{eq:pot}
\eeqn
and projecting the evolution to the torus $S^{2p+2}$:
\beqn
e^{i g_{t+1,k+1}} = 
\frac{k + e^{i(g_{t,k} - g_{t,k+2})}}
{1 + k e^{i(g_{t,k} - g_{t,k+2})}} e^{i g_{t-1,k+1}}. 
\label{eq:DSG} \eeqn
 then
with the above  redefinition the vertex equation (\ref{eq:evgt}) turns
 into the famous
Hirota equation \cite{H}, while the face equations are commonly known
as doubly discrete sine-Gordon equations (see e.g.\cite{BKP,BBR,FV2}).
Without such a redefinition along the diagonals,
but still with the  special potential given in (\ref{eq:pot}),
 the above model is related to the doubly discrete
mKdV model \cite{CN}.

Due to its relation to the sine-Gordon model (see e.g. \cite{IK})
, the to the
torus projected vertex, edge and 
face equations with the potential (\ref{eq:pot}) will be shortly
called of sine-Gordon type.
In the forthcoming we will study evolutions of  sine-Gordon type.

\section{Quantization of the models}
\subsection{General outline}
Fix an initial Cauchy path as e.g. ${\cal C}_{2T}$
(see fig. 1).

\noindent Our quantization scheme follows the common procedure to
substitute the canonical variables 
$ I_{2T} := \bigl \{(e^{ig_{2T-1,2k+1}})_{k \in \{-1,\ldots p-1\}}
\{(e^{ig_{2T,2k}})_{k \in \{0,\ldots p\}}
\bigr\}$ as functions on phase space by unitary operators 
${\bf I}_{2T+1}:= $   $
\bigl \{(G_{2T+1,2k+1})_{k \in \{-1,\ldots p-1\}}\in U({\cal H}),$  $
(G_{2T,2k})_{k \in \{0,\ldots p\}}\in U({\cal H}) \bigl \}$
obeying Weyl commutation relations,
i.e. we search for a bijection $Q: {\cal F(P)} \rightarrow U({\cal H}),$ 
$Q(e^{i g_{t,k}}) =: G_{t,k}$ with the properties
\beqnao
Q(const) &=& const \, \unity\\
Q(e^{i g_{t,k}} e^{i g_{\tilde{t},j}}) &=& Q(e^{i g_{t,k}}) Q(e^{i g_{\tilde{t},j}})\, e^{i\, phase}
\eeqnao
such that 
\begin{equation}
\begin{array}{ccll}
 [G_{t,i}, G_{t,k}] &=& 0 , &\mbox{ if $i-k$ even}\\[1mm]
 G_{t,i} G_{\tilde{t},k} &=& q^{-\frac{m}{2}}  G_{\tilde{t},k} G_{t,i},  &\mbox{ if $i-k$ odd, $i<k,\; |i-k|<2p$},\\[1mm]
 [G_{t,i}, {M^{(k)}}_t]& =& 0 &\mbox{ ($i-k$ odd}), \\[1mm]
 G_{t,i} {M^{(k)}}_{\tilde{t}} &=& q^{-m} M^{(k)}_t  G_{t,i}&\mbox{ ($i-k$ even),}\\[1mm]
 [{M^{(1)}}_t,{M^{(2)}}_{\tilde{t}}] &=& 0
\end{array}
\label{def:commu} \end{equation}
where $|t-\tilde{t}| = 1 $, $t,\tilde{t} \in \{2T,2T-1\},$ 
$q  \in S^1 \subset \KOMPLEX, $ $ m \in \NATUR$,
$$
M^{(1)} \dgleich G_{2T,2p} G^{-1}_{2T,0} \qquad M^{(2)} \dgleich G_{2T-1,2p-1} G^{-1}_{2T-1,-1}
\qquad G_{t,k+2p} \dgleich M^{(i)} G_{t,k}, 
$$
$\unity = ( G_{t,k})^0$ is the identity in $U({\cal H})$ and the product in 
$U({\cal H})$ is given by the composition of operators.
For our purpose q will always be chosen as a root of unity, i.e.
$q = e^{\frac{2 \pi i}{N}} \in S^1 \subset \KOMPLEX.$

\noindent Let ${\cal A}(G_{2T})$ be the algebra of Laurent polynomials in the
generators $ I_{2T} =$  \\    
 $ \bigl\{(G_{2T,2k})_{k \in \{0,\ldots p\}},$
$(G_{2T-1,2k+1})_{k \in \{-1,\ldots p-1\}}
\bigr\}$ ($T \in \GANZ$ fixed). 
Note that up to now the "quantization" map $Q$ shall be only defined
for the canonical variables $e^{i g_{t,k}}, t \in \{2T,2T-1\}$ and, 
modulo a phase factor,
on products of these. It shall not be specified for other functions 
on  phase space ${\cal P}$ like for example on the time one evolved variables:
$$g_{2T+1,2k-1}= g_{2T+1,2k-1}(g_{2T,2k-2},g_{2T,2k},g_{2T-1,2k-1}).$$
The idea is that we will implicitly define a quantization for these functions by 
defining an  automorphism ${\bf E}_{t,k-1}:{\cal A}(G_{2T})\rightarrow {\cal A}(G_{2T})$, 
such that :
$$
Q(e^{i g_{t+1,k-1}}) \dgleich {\bf E}_{t,k-1}(G_{t-1,k-1}) = 
{\bf E}_{t,k-1}(Q(e^{i g_{t-1,k-1}})),
$$
where $e^{i g_{t+1,k}}$is given by the classical evolution.

\noindent The automorphism ${\bf E}_{t,k-1}$ will be very much adapted to
our specific model.

We will neither discuss wether and how it would in general be possible to find such
an automorphism nor will we be concerned with a discussion of the above 
quantization procedure with respect to completeness (when e.g. extending by
linearity), uniqueness, connection
to other quantizatons etc.

\noindent Let us proceed with an explicit construction of  ${\bf E}_{t,k-1}$. 
\noindent In accordance with the classical definition (\ref{def:x})
define edge operators ($k-t$ even):
\beqna
X_{t,k} &\dgleich& G_{t,k}G_{t-1,k-1}\\
X_{t,k-1} &\dgleich& G_{t-1,k-1}G_{t,k-2}\\
 \Rightarrow M_X &\dgleich& X_{t,2p} X_{t,0}^{-1} = 
q^{-{m}} M^{(1)} M^{(2)}\label{eq:deg}\eeqna

\noindent Let ${\cal A}(X_{2T})$ be the algebra of 
Laurent polynomials  in the generators \\
$I_{2T}^{X}= $ 
$(X_{2T,k})_{k \in \{0..2p\}}.$
Define face operators 
\beqn 
P_{t,k-1} := q^{-\frac{m}{2}} G^{-1}_{t,k} G_{t,k-2} 
\label{eq:deep} \eeqn
\noindent Note that $P_{t,k-1} = X^{-1}_{t,k} X_{t,k-1} =
 X_{t+1,k}^{-1} X_{t+1,k-1}^{-1} $.

\noindent Let ${\cal A}(P_{2T})$ be the algebra of 
Laurent polynomials  in the generators \\
$I_{2T}^{P}=$ 
$\{(P_{2T-1,2k})_{k \in \{0..p-1\}}$, $(P_{2T,2k+1})_{k \in \{0..p-1\}}\}.$

\noindent Clearly the above construction can also be done for the
Cauchypath ${\cal C}_{2T-1}$, therefore the distinction between
even and odd times will be sometimes omitted.

\subsection{Almost hamiltonian quantum evolution}

\noindent Let $R_k(P_{t,n-1}) $
 be a nonvanishing Laurent polynomial in 
the face operator $P_{t,n-1}:= q^{-\frac{m}{2}} G^{-1}_{t,n} G_{t,n-2}$,
i.e. $R_k(P_{t,n-1}) \in  {\cal A}(G_{t})$ which depends on a parameter
$k \in [0,1)$. Let 
$e^{i \xi_k(P_{t,n-1}^B)} \in S^1 \subset \KOMPLEX$ be a number which depends
on a central element  $P_{t,n-1}^B \in {\cal A}(G_t)$ 
(also called Casimir) and the same parameter $k \in [0,1)$, 
where $B \in \NATUR$.
\noindent Define recursively ($n-t$ even) 
\beqna
G_{t+1,n-1} \dgleich {\bf E}_{t,n-1}(G_{t-1,n-1})&\dgleich& R_k(P_{t,n-1}) 
G_{t-1,n-1} R_k(P_{t,n-1})^{-1} e^{i \xi_k(P_{t,n-1}^B)}\nonumber\\
&=& \frac{ R_k(P_{t,n-1})}{ R_k(q^{m} P_{t,n-1})} G_{t-1,n-1} 
e^{i \xi_k(P_{t,n-1}^B)} 
\label{eq:evG}\eeqna
\beqn
{\bf E}_{t,n-1}(G_{\tilde{t},j}) \dgleich G_{\tilde{t},j} 
\quad \mbox{for}\, j \neq n-1  \, mod\, 2p; \quad \tilde{t} \in \{t,t-1\}
\label{eq:Ctrivial}
\eeqn
i.e. the automorphisms ${\bf E}_{t,n-1}$ act nontrivially only on operators
which are associated to the $(t,n-1)$'th face of the corresponding 
Cauchyzig zag ${\cal C}_t.$ 

\begin{center}
\begin{picture}(0,0)%
\epsfig{file=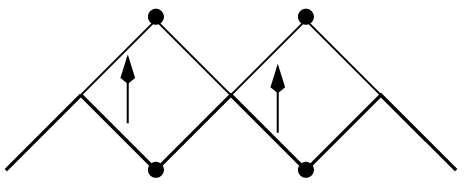}%
\end{picture}%
\setlength{\unitlength}{0.00083300in}%
\begingroup\makeatletter\ifx\SetFigFont\undefined
\def\x#1#2#3#4#5#6#7\relax{\def\x{#1#2#3#4#5#6}}%
\expandafter\x\fmtname xxxxxx\relax \def\y{splain}%
\ifx\x\y   
\gdef\SetFigFont#1#2#3{%
  \ifnum #1<17\tiny\else \ifnum #1<20\small\else
  \ifnum #1<24\normalsize\else \ifnum #1<29\large\else
  \ifnum #1<34\Large\else \ifnum #1<41\LARGE\else
     \huge\fi\fi\fi\fi\fi\fi
  \csname #3\endcsname}%
\else
\gdef\SetFigFont#1#2#3{\begingroup
  \count@#1\relax \ifnum 25<\count@\count@25\fi
  \def\x{\endgroup\@setsize\SetFigFont{#2pt}}%
  \expandafter\x
    \csname \romannumeral\the\count@ pt\expandafter\endcsname
    \csname @\romannumeral\the\count@ pt\endcsname
  \csname #3\endcsname}%
\fi
\fi\endgroup
\begin{picture}(2205,1097)(324,-4464)
\put(2506,-4171){\makebox(0,0)[lb]{\smash{\SetFigFont{12}{14.4}{rm}${\cal C}$}}}
\put(976,-3452){\makebox(0,0)[lb]{\smash{\SetFigFont{7}{8.4}{rm}$G_{t+1,n-1}$}}}
\put(1681,-3451){\makebox(0,0)[lb]{\smash{\SetFigFont{7}{8.4}{rm}$G_{t+1,n+1}$}}}
\put(1666,-4441){\makebox(0,0)[lb]{\smash{\SetFigFont{7}{8.4}{rm}$G_{t-1,n+1}$}}}
\put(976,-4443){\makebox(0,0)[lb]{\smash{\SetFigFont{7}{8.4}{rm}$G_{t-1,n-1}$}}}
\put(1666,-3947){\makebox(0,0)[lb]{\smash{\SetFigFont{7}{8.4}{rm}${\bf E}_{t,n+1}$}}}
\put(931,-3961){\makebox(0,0)[lb]{\smash{\SetFigFont{7}{8.4}{rm}${\bf E}_{t,n-1}$}}}
\end{picture}
\end{center}

\noindent Define 
$$
{\bf E}_{2t} \dgleich \prod_{n=0}^{p-1} {\bf E}_{2t,2n+1} \qquad 
{\bf E}_{2t-1} \dgleich \prod_{n=0}^{p-1} {\bf E}_{2t-1,2n}, 
$$
which is well defined since the corresponding
automorphisms ${\bf E}_{t,n-1}$ commute.
${\bf E}_t$ evolves all operators
associated to a Cauchypath ${\cal C}_t$ one time step further and
 by the definition of the evolution automorphism:
$$
{\bf E}_t(G_{t-1,n-1}) = E_{t,n-1}(G_{t-1,n-1}).
$$
Using the commutation relations in (\ref{def:commu}) and the
periodicity of the face operators, it follows immediately that the above
 automorphisms preserves
the monodromies. We will call such  automorphisms
 {\em almost hamiltonian evolution automorphisms}.

The induced evolution 
on the subalgebras
 ${\cal A}(P_T) \subset {\cal A}(X_T) \subset {\cal A}(G_T)$ reads:
\beqna
X_{t+1,n} &\dgleich& G_{t,n} G_{t+1,n-1} = G_{t,n} R_k(P_{t,n-1}) G_{t-1,n-1}
R_k(P_{t,n-1})^{-1} e^{i \xi_k(P_{t,n-1}^B)}\\
 &= & R_k(P_{t,n-1}) X_{t,n} R_k(P_{t,n-1})^{-1} e^{i \xi_k(P_{t,n-1}^B)}
=\frac{ R_k(P_{t,n-1})}{ R_k(q^{m} P_{t,n-1})} X_{t,n} e^{i \xi_k(P_{t,n-1}^B)}\nonumber \\
X_{t+1,n-1} &\dgleich& G_{t+1,n-1} G_{t,n-2} = R_k(P_{t,n-1}) G_{t-1,n-1}
R_k(P_{t,n-1})^{-1}  G_{t,n-2} e^{i \xi_k(P_{t,n-1}^B)}\nonumber \\
&=&  R_k(P_{t,n-1}) X_{t,n-1} R_k(P_{t,n-1})^{-1} e^{i \xi_k(P_{t,n-1}^B)}
=\frac{ R_k(P_{t,n-1})}{ R_k(q^{m} P_{t,n-1})} X_{t,n-1} e^{i \xi_k(P_{t,n-1}^B)} \nonumber\\
P_{t+1,n} &\dgleich& q^{-\frac{m}{2}} G^{-1}_{t+1,n+1} G_{t+1,n-1} \\
&=& q^{-\frac{m}{2}}
R_k(P_{t,n+1}) G_{t-1,n+1}^{-1} R_k(P_{t,n+1})^{-1}  e^{-i \xi_k(P_{t,n+1}^B)}\\
&&R_k(P_{t,n-1})G_{t-1,n-1}R_k(P_{t,n-1})^{-1}
 e^{i \xi_k(P_{t,n-1}^B)} \nonumber \\
&=& \frac{ R_k(P_{t,n-1})}{ R_k(q^{m} P_{t,n-1})} 
\frac{ R_k(P_{t,n+1})}{ R_k(q^{-m} P_{t,n+1})} P_{t-1,n}
e^{i \xi_k(P_{t,n-1}^B)} e^{-i \xi_k(P_{t,n+1}^B)}\nonumber \\
&=:&  {\bf E}_{t,n+1}{\bf E}_{t,n-1} (P_{t-1,n}) 
\eeqna
\noindent ($n-t$ even).

Now $I^P_{2T} = \{(P_{2T-1,2n})_{n \in \{0..p-1\}}$, $(P_{2T,2n+1})_{n \in \{0..p-1\}}\}$
or  $ I^P_{2T-1},$ respectively,  shortly denoted by $I_T^P,$ 
is an initial configuration of arbitrary unitary periodic (face) operators, i.e. 
$P_{t,2p+n}= P_{t,n} \in {\cal A}(P_{T}),\, n \in \GANZ$
which
 obey  the commutation rules:
$$
[P_{t,n},P_{t,j}] = 0 \quad n \neq j
$$
$$
[P_{2T \pm 1,n}, P_{2T,n + j}] = 0 \qquad \mbox{for} \quad j  \in \{2...2p-2\}
\quad mod \, 2p
$$
\beqn
P_{t,n} P_{\tilde{t},n+1} = q^{-m} P_{\tilde{t},n+1} P_{t,n}
\label{def:Pcomm}\eeqn
\noindent $t,\tilde{t} \in \{2T, 2T \pm 1\}$,$|t-\tilde{t}|=1$

\noindent Consider the  automorphism ${\bf E}_{t,n+1}{\bf E}_{t,n-1}
:{\cal A}(P_T)\rightarrow {\cal A}(P_T)$ which was recursively defined by:
\beqna
&&{\bf E}_{t,n+1}{\bf E}_{t,n-1} (P_{t-1,k}) \\
&\dgleich&
R_k(P_{t,n+1}) R_k(P_{t,n-1}) P_{t-1,n} R_k(P_{t,n-1})^{-1} R_k(P_{t,n+1})^{-1} 
e^{i \xi_k(P_{t,n-1}^B)} e^{-i \xi_k(P_{t,n+1}^B)} \nonumber\label{eq:deP}
\eeqna
and ${\bf E}_{t,n+1}{\bf E}_{t,n-1}$ acting trivially on all other faces 
along the corresponding Cauchy zig zag ${\cal C}_t$.
Since
$$
P_{t+1,n} \dgleich {\bf E}_{t,n+1}{\bf E}_{t,n-1} (P_{t-1,n}). 
$$
it follows by the  definition (\ref{eq:deP}) that :
\beqna
&&R_k(P_{t+1,n}) \nonumber \\
&=& R_k(R(P_{t,n+1}) R_k(P_{t,n-1}) P_{t-1,n}  R_k(P_{t,n+1})^{-1} R_k(P_{t,n+1})^{-1}
e^{i \xi_k(P_{t,n-1}^B)} e^{-i \xi_k(P_{t,n+1}^B)})\nonumber \\
&=& R_k(P_{t,n+1}) R_k(P_{t,n-1}) R_k(e^{i \xi_k(P_{t,n-1}^B)} 
e^{-i \xi_k(P_{t,n+1}^B)} P_{t-1,n}) R_k(P_{t,n-1})^{-1} R_k(P_{t,n+1})^{-1}.\nonumber\\
&&
\label{eq:Conj} \eeqna
Define the operators ($t \in \GANZ$)
$$
{\bf R}_{2t-1} := \prod_{n=0}^{p-1} R_k(P_{2t-1,2n}) 
\qquad {\bf R}_{2t} := \prod_{n=0}^{p-1} R_k(P_{2t,2n+1}).
$$
\begin{Proposition}
The operators ${\bf R}_{t} \in {\cal A}(P)$ defined as above evolve as:
\beqnao
{\bf R}_{2t+1} &=& {\bf R}_{2t}
\prod_{n=0}^{p-1}  R_k(e^{i \xi_k(P_{2t,2n-1}^B)} e^{-i \xi_k(P_{2t,2n+1}^B)}
 P_{2t-1,2n})
 {\bf R}_{2t}^{-1} \\
{\bf R}_{2t} &=& {\bf R}_{2t-1}
\prod_{n=0}^{p-1}  R_k(e^{i \xi_k(P_{2t-1,2n}^B)} e^{-i \xi_k(P_{2t-1,2n+2}^B)} 
P_{2t-2,2n+1})
{\bf R}_{2t-1}^{-1} 
\eeqnao
\label{theor:Rkeep}
\end{Proposition}

\noindent {\bf Proof:} Using the commutation relations in (\ref{def:Pcomm}) and (\ref{eq:Conj})
and the periodicity of the face operators the proof is straightforward.
\noindent $\Box$

\begin{Corollary}
If $e^{i \xi_k(P_{t,n-1}^B)} e^{-i \xi_k(P_{t,n+1}^B)}= \unity$ f.a.
$t,x \in \GANZ$ ($k-t$ even) then 
${\bf R}_{t+1} {\bf R}_{t} = {\bf R}_{t} {\bf R}_{t-1}$ is constant.
\label{cor:IntR} \end{Corollary}
Since in this case the evolution for the face variables is given by conjugation
$$
 P_{t+1,n} = {\bf R}_t {\bf R}_{t-1} P_{t-1,n} {\bf R}_{t-1}^{-1} {\bf R}_t^{-1}
$$
 the operator ${\bf R}_{t} {\bf R}_{t-1}$ can be viewed as the
discrete analog of a continous hamiltonian time evolution, i.e.
$$
{\bf R}_{t} {\bf R}_{t-1} \leftrightarrow e^{i H \Delta t_0}
\qquad \Delta t_0 \, \mbox{fixed}.
$$
This should justify, why we called the automorphism constructed in 
(\ref{eq:evG}),(\ref{eq:Ctrivial}) {\em almost} hamiltonian.

\section{The quantum discrete sine-Gordon model}
\subsection{Constructing the evolution automorphism}
\begin{Theorem}
\label{theor:Rmatrix}
Let $\hat{x}$ be an element of a $*$-algebra over $\KOMPLEX$, such that $x^{-1} = x^*$
and $\hat{x}^B := \underbrace{\hat{x}  \cdot \ldots \cdot \hat{x}}_{B \, \, times}$
is a multiple of the identity element $\unity$ in the algebra, i.e.
$\hat{x}^B = x^B \unity$, with $x^B \in S^1 \subset \KOMPLEX$.
Let $k \in [0,1), B = \frac{N}{gcd(m,N)}, q = e^{\frac{2 \pi i}{N}}$.
Choose any root 
\beqn
e^{i \xi_k} = e^{i \xi_k(\hat{x}^B)} := \left(\frac{1+  k^B x^B (-1)^{B-1}}{k^B +  
x^B (-1)^{B-1}}\right)^{\frac{1}{B}} \unity \quad \in S^1 \unity.
\label{eq:dxi}
\eeqn
Define $R_k(\hat{x}) := \sum_{j=0}^{B-1} l_j \hat{x}^j$ with
$$
l_j := \prod_{n=1}^{j} \frac{e^{i \xi_k} q^{m(n-1)}-k}{1-e^{i \xi_k} k q^{mn}}
$$
then  $R_k(\hat{x})$ satisfies the functional equation:
\beqn
R_k(\hat{x}) = \frac{k + \hat{x}}{1 + k \hat{x}}\,R_k(q^m \hat{x})
e^{i \xi_k}.
\label{eq:wich}\eeqn
\end{Theorem}

\noindent {\bf Proof:}
By straightforward verification.

\noindent $\Box$

\noindent The above theorem is a generalization of the results found
in \cite{V} and \cite{FZ}, where it was assumed that
$\hat{x}^B = \unity$. As it turns out a generalization 
to general Casimirs $\hat{x}^B$ is important for obtaining
an evolution with nontrivial classical background as 
first described in \cite{BBR}, where a solution for the 
above functional equation (\ref{eq:wich})was indicated for the case
$m=2, N=$ odd. Solutions to the above equation for 
$q^m$ not being a root of unity can also be found in \cite{BBR}.

Unfortunately, despite the suggestive notation the numbers
$ e^{i \xi_k(\hat{x}^B)}$ depend not only on the Casimirs
$\hat{x}^B$ and the real constant $k \in [0,1)$, but also on the choice 
of a $B$'th root.
Clearly once a choice is made (for all times) one can use the operators 
$R_k(\hat{x})$ together with the chosen roots (now viewed as
 functions in the Casimirs) for defining an almost
hamiltonian quantum evolution for the sine-Gordon model as in
(\ref{eq:evG}).

However the fixing of the roots $e^{i \xi_k(\hat{x}^B)}$  for
all times contradicts the idea of defining an evolution, by
a local process.
In the following we want to show that given an initial choice of roots,
it is possible to define an unique evolution for the above roots.
Nevertheless in order to accomplish this task
we need to extend the algebra ${\cal A}(X_T)$ 
by the central elements $ e^{i \xi_k(P_T^B)}$ and will denote this new 
algebra by ${\cal A}(X_T)^e.$

\subsection{Light cone shifts}
The doubly discrete sine-Gordon equation, as well as the above described 
equations of sine-Gordon type (see e.g. (\ref{eq:DSG})) are,
as in the continous case, invariant under light cone shifts, i.e. if
$g_{t,k}: \GANZ^2 \rightarrow \REELL$ is a solution to (\ref{eq:evgt}) then
$g_{t \pm n,k \pm n}$ is also a solution. In this sense space time shifts
can be lifted to automorphisms on covariant phase space and
can be interpreted as symplectomorphisms  \cite{EK}.

In the previous section we found a quantization of another 
(yet trivial) symplectomorphism (\cite{AM, GS}) on phase space, namely 
time evolution. It would be now only consequent to find quantized
analogs of the above mentioned light cone shifts. This will be done
by defining quantized space translations of half the lattice spacing distance
and then applying the time automorphism. Since
translations of half the lattice spacing distance are hard to define
on the vertex operators, as one would have to go over to
the dual lattice, one has to restrict oneself to the
edge algebra ${\cal A}(X_T).$ 

For constructing the above mentioned space shifts, 
we will follow an idea developped in \cite{FV3}, where such space  shifts 
were suggested for the case of a special choice of vertex  monodromies. As it
will turn out the treatment of the more general case 
will result in a possibility to fix the above roots in
a very natural way.

\begin{Lemma} The quotient of the two vertex monodromies
\beqnao
M^{(1)} (M^{(2)})^{-1}&=&
P_{2t\pm 1,0} P_{2t \pm 1,2} \ldots P_{2t\pm 1,2p-2}
P_{2t,2p-1}^{-1} P_{2t,2p-3}^{-1} \ldots P_{2t,1}^{-1}\\
&=& q^{-2 m p +m } X^2_{2t,0} X^{-2}_{2t,1} X^2_{2t,2} \ldots X^2_{2t,2p-1} M_X \\
&=&  q^{-2 m p +m } X^2_{2t+1,0} X^{-2}_{2t+1,1} X^2_{2t+1,2} \ldots X^2_{2t+1,2p-1} M_X 
\eeqnao
is a Casimir in ${\cal A}(X_{T})$.
\end{Lemma}
\begin{Demand} For establishing quantum space time shifts demand 
\bbl{a.)}
\item[a.)]The Casimirs 
$$P_{2t-1,0} P_{2t- 1,2} \ldots P_{2t-1,2p-2}
P_{2t,2p-1}^{-1} P_{2t,2p-3}^{-1} \ldots P_{2t,1}^{-1}
$$
and $P_{t,k-1}^B$  ($B$ as before)
shall be multiples of the identity within ${\cal A}(X_{T})$.
\item[b.)] The roots $e^{i \xi_0(P^B_{t,k-1})} = 
(P_{t,k-1} (-1)^{B-1})^{-\frac{1}{B}}$ (see \ref{eq:dxi})
shall be fixed in such a way that
$$
\prod_{k=0}^{p-1}(P_{2t-1,2k}^B (-1)^{B-1})^{\frac{1}{B}} 
\prod_{k=0}^{p-1}(P_{2t,2k-1}^B (-1)^{B-1})^{-\frac{1}{B}} 
= \prod_{k=0}^{p-1}P_{2t-1,2k} 
\prod_{k=0}^{p-1}P_{2t,2k-1}^{-1} 
$$\ebl
By the definition of an almost hamiltonian quantum evolution
demand a.) holds automatically for all times $t$, if it is true 
at an initial time $T$. The same is valid for demand b.) which will become
evident soon, therefore the index $t$ in the above demands 
refers to all times.
\label{dem:Cas} \end{Demand}
Define
\beqn
S_t^{-1} \dgleich \prod_{k=2}^{2p\,\leftarrow} R_0 (X_{t,k}^{-1} X_{t,k-1}) =
R_0 (X_{t,2p}^{-1} X_{t,2p-1}) R_0 (X_{t,2p-1}^{-1} X_{t,2p-2}) \ldots
R_0 (X_{t,2}^{-1} X_{t,1})
\eeqn
\begin{Proposition}
For all $k \in \GANZ$  
$$
S_t^{-1} X_{t,k} S_t = q^{-m} e^{i \xi_0((X_{t,k}^{-1} X_{t,k-1})^B)} X_{t,k-1}
$$
where $e^{i \xi_0(x^B)} = (x^B (-1)^{B-1})^{-\frac{1}{B}}$ as in (\ref{eq:dxi}).
\end{Proposition}

\noindent {\bf Proof:}

\noindent If $n \in \{2 \ldots 2p \}$ then by the commutation rules of the 
edge variables
\beqnao
&&S_t^{-1} X_{t,n} S_t =\\ 
&=&\prod_{k=n}^{2p\,\leftarrow} R_0 (X_{t,k}^{-1} X_{t,k-1})
X_{t,n} \prod_{k=n}^{2p\,\rightarrow} R_0 (X_{t,k}^{-1} X_{t,k-1})^{-1}\\
&=& \prod_{k=n+1}^{2p\,\leftarrow} R_0 (X_{t,k}^{-1} X_{t,k-1})
\frac{R_0(X_{t,n}^{-1} X_{t,n-1})}{R_0(q^{m} X_{t,n}^{-1} X_{t,n-1})}
 X_{t,n} \prod_{k=n+1}^{2p\,\rightarrow} R_0 (X_{t,k}^{-1} X_{t,k-1})^{-1}\\
&=& \prod_{k=n+1}^{2p\,\leftarrow} R_0 (X_{t,k}^{-1} X_{t,k-1})
X_{t,n}^{-1} X_{t,n-1} X_{t,n}  
\prod_{k=n+1}^{2p\,\rightarrow} R_0 (X_{t,k}^{-1} X_{t,k-1})^{-1}
 e^{i \xi_0((X_{t,n}^{-1} X_{t,n-1})^B)}\\
&=&q^{-m}\prod_{k=n+1}^{2p\,\leftarrow} R_0 (X_{t,k}^{-1} X_{t,k-1}) 
 X_{t,n-1}
\prod_{k=n+1}^{2p\,\rightarrow} R_0 (X_{t,k}^{-1} X_{t,k-1})^{-1}
 e^{i \xi_0((X_{t,n}^{-1} X_{t,n-1})^B)}\\
&=& q^{-m} e^{i \xi_0((X_{t,n}^{-1} X_{t,n-1})^B)} X_{t,n-1}
\eeqnao
Analogously one obtains
\beqna
&&S_t^{-1} X_{t,1} S_t = \nonumber \\ 
&=& \prod_{k=3}^{2p\,\leftarrow} R_0 (X_{t,k}^{-1} X_{t,k-1})
X_{t,2}^{-1} X_{t,1}^2 \prod_{k=3}^{2p\,\rightarrow} R_0 (X_{t,k}^{-1}
 X_{t,k-1})^{-1}
 e^{i \xi_0((X_{t,2}^{-1} X_{t,1})^B)} \nonumber\\
&=& \prod_{k=4}^{2p\,\leftarrow} R_0 (X_{t,k}^{-1} X_{t,k-1})
\frac{R_0(X_{t,3}^{-1} X_{t,2})}{R_0(q^{-m} X_{t,3}^{-1} X_{t,2})}
X_{t,2}^{-1} X_{t,1}^2  \prod_{k=4}^{2p\,\rightarrow} R_0 
(X_{t,k}^{-1} X_{t,k-1})
e^{i \xi_0((X_{t,2}^{-1} X_{t,1})^B)}\nonumber\\
&=& \prod_{k=4}^{2p\,\leftarrow} R_0 (X_{t,k}^{-1} X_{t,k-1})
q^{2m} X_{t,3} X_{t,2}^{-2} X_{t,1}^2
 \prod_{k=4}^{2p\,\rightarrow} R_0 (X_{t,k}^{-1} X_{t,k-1})
e^{-i \xi_0((X_{t,3}^{-1} X_{t,2})^B)+ 
i \xi_0((X_{t,2}^{-1} X_{t,1})^B)} \nonumber\\
&=&
q^{2m(p-1)} X_{t,2p}^{-1} \prod_{k=1}^{2p-1} X_{t,k}^{2(-1)^{k+1}}
\prod_{k=1}^{2p}  e^{(-1)^{k}i \xi_0((X_{t,k}^{-1} X_{t,k-1})^B)}
\, e^{i \xi_0((X_{t,1}^{-1} X_{t,0})^B)} \nonumber\\
&\stackrel{b.)}{=}& q^{-m} e^{i \xi_0((X_{t,1}^{-1} X_{t,0})^B)} X_{t,0}
\label{eq:MonX}\eeqna
if we suppose that demand b.) holds. With $S_t^{-1} M_X S_t = M_X$ the assertion follows.

\noindent $\Box$

\noindent It is easy to show that:
$$
(S_t^{(l)})^{-1} X_{t,n} S_t^{(l)} = q^{-m} e^{i \xi_0((X_{t,n}^{-1} X_{t,n-1})^B)} 
X_{t,n-1}
$$
where
$$
(S_t^{(l)})^{-1} \dgleich \prod_{k=2+l}^{2p+l\,\leftarrow} R_0 (X_{t,k}^{-1} X_{t,k-1})
$$
Therefore the index $l$ is irrelevant and will be skipped, also if 
$(S_t^{(l)})^{-1}$ instead of $(S_t^{(0)})^{-1}$ is used.
Clearly one can also define an automorphism of the above kind
by redefining
$$
\hat{S}_t^{-1}\dgleich \prod_{k=2+l}^{2p+l\,\leftarrow} R_0 (\alpha X_{t,k}^{-1} X_{t,k-1})
\qquad \alpha \in S^1,
$$
which will act as:
$$
\hat{S}_t^{-1} X_{t,n} \hat{S}_t = \alpha q^{-m} e^{i \xi_0((X_{t,n}^{-1} X_{t,n-1})^B)} 
X_{t,n-1},
$$
a fact to be used later.

\noindent The automorphism ${\bf S}_t^{-1}: {\cal A}(X_T) \rightarrow {\cal A}(X_T)$
defined on the operators $X_{t,n}$ as
$${\bf S}_t^{-1} (X_{t,n}) := 
q^m e^{-i \xi_0((X_{t,n}^{-1} X_{t,n-1})^B)} S_t^{-1} X_{t,n} S_t
= X_{t,n-1}
$$
can be interpreted as  a shift of these edge operators in space direction.
As  an automorphism on ${\cal A}(P_T)$, 
the picture of the action of ${\bf S}_t^{-1}$ is  a little different, we find
\beqnao
S_t^{-1} P_{t,k-1} S_t &=& 
P_{t-1,k-2}\; e^{-i \xi_0(P_{t,k-1}^B)} + e^{i \xi_0(P_{t-1,k-2}^B)}\\
S_t^{-1} P_{t-1,k-2} S_t &=& 
P_{t,k-3}\;  e^{-i \xi_0(P_{t,k-2}^B)} + e^{i \xi_0(P_{t-1,k-3}^B)}
\eeqnao
Hence ${\bf S}_t$ applied to face operators is rather an
up and down  shift in lightcone direction than a shift in space:

\begin{center}
\begin{picture}(0,0)%
\epsfig{file=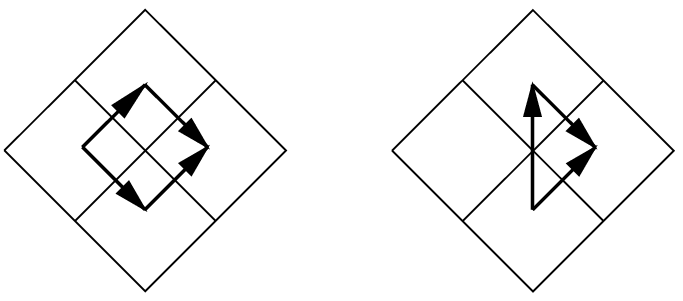}%
\end{picture}%
\setlength{\unitlength}{0.00083300in}%
\begingroup\makeatletter\ifx\SetFigFont\undefined
\def\x#1#2#3#4#5#6#7\relax{\def\x{#1#2#3#4#5#6}}%
\expandafter\x\fmtname xxxxxx\relax \def\y{splain}%
\ifx\x\y   
\gdef\SetFigFont#1#2#3{%
  \ifnum #1<17\tiny\else \ifnum #1<20\small\else
  \ifnum #1<24\normalsize\else \ifnum #1<29\large\else
  \ifnum #1<34\Large\else \ifnum #1<41\LARGE\else
     \huge\fi\fi\fi\fi\fi\fi
  \csname #3\endcsname}%
\else
\gdef\SetFigFont#1#2#3{\begingroup
  \count@#1\relax \ifnum 25<\count@\count@25\fi
  \def\x{\endgroup\@setsize\SetFigFont{#2pt}}%
  \expandafter\x
    \csname \romannumeral\the\count@ pt\expandafter\endcsname
    \csname @\romannumeral\the\count@ pt\endcsname
  \csname #3\endcsname}%
\fi
\fi\endgroup
\begin{picture}(3237,1697)(1310,-4446)
\put(3961,-3736){\makebox(0,0)[lb]{\smash{\SetFigFont{7}{8.4}{rm}${\bf S}_{T}$}}}
\put(3677,-4425){\makebox(0,0)[lb]{\smash{\SetFigFont{8}{9.6}{rm}Figure 2.2}}}
\put(1757,-4410){\makebox(0,0)[lb]{\smash{\SetFigFont{8}{9.6}{rm}Figure 2.1}}}
\put(4113,-3301){\makebox(0,0)[lb]{\smash{\SetFigFont{7}{8.4}{rm}${\bf S}_{T+1}$}}}
\put(3693,-3042){\makebox(0,0)[lb]{\smash{\SetFigFont{6}{7.2}{rm}$P_{T+1,k}$}}}
\put(1801,-3056){\makebox(0,0)[lb]{\smash{\SetFigFont{6}{7.2}{rm}$P_{T+1,k}$}}}
\put(2087,-3720){\makebox(0,0)[lb]{\smash{\SetFigFont{7}{8.4}{rm}${\bf S}_{T}$}}}
\put(3663,-3826){\makebox(0,0)[lb]{\smash{\SetFigFont{6}{7.2}{rm}$P_{T-1,k}$}}}
\put(1817,-3855){\makebox(0,0)[lb]{\smash{\SetFigFont{6}{7.2}{rm}$P_{T-1,k}$}}}
\put(2222,-3271){\makebox(0,0)[lb]{\smash{\SetFigFont{7}{8.4}{rm}${\bf S}_{T+1}$}}}
\put(2297,-3435){\makebox(0,0)[lb]{\smash{\SetFigFont{6}{7.2}{rm}$P_{T,k+1}$}}}
\put(1592,-3690){\makebox(0,0)[lb]{\smash{\SetFigFont{7}{8.4}{rm}${\bf S}_{T}$}}}
\put(1322,-3450){\makebox(0,0)[lb]{\smash{\SetFigFont{6}{7.2}{rm}$P_{T,k-1}$}}}
\put(1536,-3298){\makebox(0,0)[lb]{\smash{\SetFigFont{7}{8.4}{rm}${\bf S}_{T+1}$}}}
\put(4158,-3451){\makebox(0,0)[lb]{\smash{\SetFigFont{6}{7.2}{rm}$P_{T,k+1}$}}}
\put(3571,-3481){\makebox(0,0)[lb]{\smash{\SetFigFont{8}{9.6}{rm}${\bf E}_t$}}}
\end{picture}
\end{center}
 
\noindent Fix an initial time e.g. $T=$ odd. Then by definition the operators
\beqnao
S_{T}^{-1} &=& R_0(P_{T-1,2p-1}) R_0(P_{T,2p-2}) \ldots R_0(P_{T-1,1})\\
S_{T+1}^{-1} &=& R_0(P_{T+1,2p-1}) R_0(P_{T,2p-2}) \ldots R_0(P_{T+1,1})
\eeqnao
are given by the  choice of initial operators
$\bigl \{ (P_{T-1,2k+1})_{k \in \{0 \ldots p-1\}},
(P_{T,2k})_{k \in \{0 \ldots p-1\}} \bigr \}$ and the time evolved
face operators $(P_{T+1,2k+1})_{k \in \{0 \ldots p-1\}}$. 
The operator $P_{T+1,2k+1}$ can now 
be obtained by shifting the operator $P_{T,2k}$ or by applying
the time evolution to $P_{T-1,2k+1}$ , i.e. the in figure 2.2 depictured
shifts have to commute.

\noindent By the commutation relations of the face operators it is straightforward
to find:
\begin{Lemma}
\beqnao
&&R_0(P_{T+1,2p-1}) R_0(P_{T,2p-2}) \ldots R_0(P_{T+1,1})\\
&&= {\bf R}_T R_0(e^{- i \xi_0(P_{T,2p}^B) + i \xi_0 (P_{T,2p-2}^B)}
P_{T-1,2p-1}) R_0(P_{T,2p-2}) \\
&&\ldots R_0(e^{- i \xi_0(P_{T,2}^B) + 
i \xi_0 (P_{T,0}^B)}P_{T-1,1}) {\bf R}_T^{-1}.
\eeqnao
\label{lem:comut}\end{Lemma}
Hence
\beqnao
&&S_{T+1}^{-1} P_{T,2n} S_{T+1} 
\stackrel{!}{=} P_{T+1,2n-1} e^{-i \xi_0(P_{T,2n}^B) + i \xi_0(P_{T+1,2n-1}^B)}\\
&\stackrel{\ref{lem:comut}}{=}& {\bf R}_T
R_0(e^{- i \xi_0(P_{T,2p}^B) + i \xi_0 (P_{T,2p-2}^B)}
P_{T-1,2p-1}) R_0(P_{T,2p-2}) \\
&&\ldots R_0(e^{- i \xi_0(P_{T,2}^B)  
 + i \xi_0 (P_{T,0}^B)}P_{T-1,1}) P_{T,2k} \\
&&({\bf R}_T R_0(e^{- i \xi_0(P_{T,2p}^B) + i \xi_0 (P_{T,2p-2}^B)}
P_{T-1,2p-1}) R_0(P_{T,2p-2}) \\
&&\ldots R_0(e^{- i \xi_0(P_{T,2}^B)  
 + i \xi_0 (P_{T,0}^B)}P_{T-1,1}))^{-1}\\
&=& {\bf R}_T P_{T-1,2n-1} {\bf R}_{T-1}^{-1} e^{-i \xi_0(P_{T,2n}^B) + 
i \xi_0((e^{- i \xi_k(P_{T,2n}^B) +i \xi_k(P_{T,2n-2}^B)}P_{T-1,2n-1})^B)} 
\\
&& e^{- i \xi_k(P_{T,2n}^B) +i \xi_k(P_{T,2n-2}^B)}\\
&=& P_{T+1,2n-1} e^{-i \xi_0(P_{T,2n}^B) + 
i \xi_0((e^{- i \xi_k(P_{T,2n}^B) +i \xi_k(P_{T,2n-2}^B)}P_{T-1,2n-1})^B)} 
\eeqnao
Comparing both sides of the equation, one obtains
finally the  compatibility condition:
\beqn
e^{i \xi_0(P_{T+1,2n-1}^B)} \stackrel{!}{=} 
e^{- i \xi_k(P_{T,2n}^B) +i \xi_k(P_{T,2n-2}^B)} e^{i \xi_0(P_{T-1,2n-1}^B)}
\label{eq:Rootseq} \eeqn
For defining the evolution we had already fixed the roots
on the right hand side of equation (\ref{eq:Rootseq}), so that 
equation (\ref{eq:Rootseq}) determines the roots at one
time step further.

Equation (\ref{eq:Rootseq}) fixes the roots at $k=0$. It is 
easy to see that one can stay on one leave, when extending 
to arbitrary $k \in [0,1)$. Hence also the evolution of the roots 
$e^{- i \xi_k(P)}$ has been now defined.
As a direct consequence of equation (\ref{eq:Rootseq}),
it follows that demand b.) holds
for all times since the classical monodromies are as well
as their quantum counterparts integrals of motion \cite{K}.
 
Note that if one takes the $B'th$ power of
 equation (\ref{eq:Rootseq}) then one obtains
$$
P_{T+1,2n-1}^B = \frac{k^B + P_{T,2n}^B (-1)^{B-1}}{1+k^B P_{T,2n}^B (-1)^{B-1}}
\frac{1+k^B P_{T,2n-2}^B (-1)^{B-1}}{k^B + P_{T,2n-2}^B (-1)^{B-1}} P_{T-1,2n-1}^B.
$$
Hence the "classical" variables $P^B_{t,k}$ satisfy also an equation of 
sine-Gordon type. A fact which was first noticed in \cite{BBR} by using
the commutation relations of the face operators and computing $P_{T+1,2k-1}^B $.

It is now straightforward to show that as an immediate consequence 
$$
{\bf S}_{T+1}(X_{T+1,k}) = {\bf E}_T \circ {\bf S}_T  \circ
{\bf E}_T^{-1}(X_{T+1,k})
$$
i.e. shifts in time and space direction commute if (\ref{eq:Rootseq})
is satisfied.

The picture of the above developped quantum evolution looks
considerably complicate. It simplifies by a great amount, if
one restricts oneself to the case of corollary (\ref{cor:IntR}):
\begin{Proposition}
Let ${\bf S}_T$ be an automorphism on ${\cal A}(X_T)$ such that 
at an initial time $T$
\beqn
{\bf R}_T = {\bf S}_T({\bf R}_{T-1}) = {\bf S}_T^{-1}({\bf R}_{T-1})
\label{eq:SR} \eeqn
where recursively 
\beqn
{\bf R}_{t+1}= {\bf R}_t {\bf R}_{t-1} {\bf R}_t^{-1}
\label{eq:SRev} 
\eeqn
then 
${\bf R}_{t} = {\bf R}_T {\bf S}_T ({\bf R}_{t-1}) {\bf R}_T^{-1} \; \mbox{for all} 
\; t \in \GANZ \geq T.$
\label{prop:Shif}\end{Proposition}

\noindent {\bf Proof:}

\noindent By (\ref{eq:SR}) and (\ref{eq:SRev})
\beqna
{\bf R}_T = {\bf R}_T {\bf S}_T ({\bf R}_{T-1}) {\bf R}_T^{-1} \label{eq:Ind0}\\
{\bf R}_{T+1} = {\bf R}_T {\bf S}_T ({\bf R}_{T}) {\bf R}_T^{-1} \label{eq:Ind1}
\eeqna
For completing the inductional argument assume that the following is true
\beqna
{\bf R}_{T+n} = {\bf R}_T {\bf S}_T ({\bf R}_{T-1+n}) {\bf R}_T^{-1} \label{eq:Ind2}\\
{\bf R}_{T+1+n} = {\bf R}_T {\bf S}_T ({\bf R}_{T+n}) {\bf R}_T^{-1} \label{eq:Ind3}
\eeqna
Hence
\beqna
{\bf R}_{T+n+2} &\stackrel{(\ref{eq:SRev})}{=}&{\bf R}_{T+n+1} {\bf R}_{T+n} {\bf R}_{T+n+1}^{-1} \nonumber\\
&\stackrel{(\ref{eq:Ind2},\ref{eq:Ind3})}{=}& 
{\bf R}_T {\bf S}_T ({\bf R}_{T+n}) {\bf R}_T^{-1}
{\bf R}_T {\bf S}_T ({\bf R}_{T-1+n}) {\bf R}_T^{-1}
({\bf R}_T {\bf S}_T ({\bf R}_{T+n}) {\bf R}_T^{-1})^{-1}\nonumber\\
&\stackrel{(\ref{eq:SRev})}{=}&
{\bf R}_T {\bf S}_T ({\bf R}_{T+n+1}) {\bf R}_T^{-1} \label{eq:Ind4}
\eeqna
analogously ${\bf R}_{T+n+3} \stackrel{(\ref{eq:Ind3},\ref{eq:Ind4})}{=}
{\bf R}_T {\bf S}_T ({\bf R}_{T+n+2}) {\bf R}_T^{-1}.$

\noindent $\Box$

\begin{Proposition}
If ${\bf S}_T$ is an automorphism such at initial time $T$
$$
X_{T,k} = {\bf S}_T(X_{T,k-1}) = {\bf S}_T^{-1}(X_{T,k+1})
\qquad \mbox{and} \quad X_{t+1,k} =  {\bf R}_t X_{t,k} {\bf R}_t^{-1}
$$
for all $t \in \NATUR$ and with ${\bf R}_t$ as in proposition 
(\ref{prop:Shif}) then
$$
X_{t+1,k} = {\bf R}_T {\bf S}_T (X_{t,k-1}){\bf R}_T^{-1}
$$
\label{prop:Shif2} \end{Proposition}

\noindent {\bf Proof:}

\noindent By induction as above and by the use of proposition
(\ref{prop:Shif}).

\noindent $\Box$

\noindent
The connection to models of statistical mechanics is now evident.
We find
\beqna
X_{t+2,k+1} &=&  {\bf R}_T {\bf S}_T^{-1}
( {\bf R}_T {\bf S}_T (X_{t,k+1})  {\bf R}_T^{-1} ) {\bf R}_T^{-1}\nonumber\\
 &=&  {\bf R}_T {\bf S}_T^{-1}({\bf R}_T) X_{t,k+1} {\bf S}_T^{-1}({\bf R}_T^{-1})
{\bf R}_T^{-1}.
\label{eq:Xcom}
\eeqna
Moreover ${\bf R}_T $ is a product of "local amplitudes" 
$R(P_{T,k-1})$ associated to the faces at time $T$ within
 the light cone lattice,
 hence ${\bf S}_T^{-1}({\bf R}_T)$ is a product of "local amplitudes" 
$R(P_{t,k-1})$ associated to faces which are shifted in lightcone
direction of the original faces. Because of (\ref{eq:Xcom}) this picture
is the same all over the lattice, hence we can interpret $R_T$ as a 
kind of transfermatrix  (though with complex weights).

Another fortunate consequence of the above is that for investigating the 
evolution it suffices to control the first time step, everything else is obtained
by applying the light cone shifts ${\bf E}_T \circ {\bf S}_T$. This
is espiacially important for the construction of integrals of motion,
since if one finds an operator $H_T$, which commutes with the above 
light cone shifts, then this will be automatically an integral of motion.

In the next section an example of such a 'static' quantum field theory will be discussed.
\section{Relations to the massive Thirring model}

Choose an initial time $T=$ odd. 

Let 
\beqn
B = \left( \ba{cc} 1&0\\0&-1 \ea \right)\quad 
S = \left( \ba{cc} 0&1\\1&0 \ea \right) \quad
C =  \left( \ba{cc} 1&0\\0&i \ea \right) 
\label{eq:B_def} \eeqn
The notation is adopted from viewing $B$ as a "{\bf B}oost" and $S$ as
a "{\bf S}hift" operator. Note that 
$$C S C^{-1}= -i B S \qquad C B S C^{-1}= -i  S.$$
$B,S,C$ are operators acting on the  Hilbertspace 
${\cal H} = \KOMPLEX^2$. Define  for any operator $A$ 
on ${\cal H}$ an operator $A_l$ on a "big"
Hilbertspace ${\cal H}^{ 2 p } =\bigotimes_{k=0}^{2p-1}\KOMPLEX^2$ by
$$
A_l = \underbrace{\unity}_{2p-1'th \, site}  \otimes \; \unity 
\ldots \underbrace{A}_{l'th \, site} 
\ldots \unity \otimes \underbrace{\unity}_{0'th \,  site} 
$$
Let 
$$
{\bf C} \dgleich \prod_{l=0}^{2p-1} C_l.
$$
Define
\beqna
X_{T,2k} &\dgleich& S_{2k} \prod_{l=0}^{2k-1\,\leftarrow} B_l
= S_{2k}B_{2k-1} \ldots B_0,\\
X_{T,2k+1}  &\dgleich& - B_{2k+1} S_{2k+1}  \prod_{l=0}^{2k\,\leftarrow}  B_l,\\
M_X &\dgleich& \unity \otimes \unity \ldots \otimes \unity 
\eeqna
where $X_{T,0} = S_0$. Denote shortly $BS_{2k+1} \dgleich B_{2k+1} S_{2k+1}.$

\noindent The above defined operators satisfy the commutation relations of the edge 
operators 
$$
X_{T,k} X_{T,k+1} = q^{-m} X_{T,k+1}X_{T,k} 
\qquad 
X_{T,k} M_X = q^{-2m} M_X X_{T,k}
$$
for $q^{-m} = -1.$ They are periodic and 
\beqn
X_{T,2k}^2 = \unity \quad X_{T,2k+1}^2 = - \unity
\label{eq:idem}\eeqn
The corresponding face operators are almost all bilocal in terms
of the tensorproduct:
$$
\ba{lllllll}
P_{T,2k} &:=& X_{T,2k+1}^{-1} X_{T,2k} &\qquad& P_{T-1,2k+1} &:=& X_{T,2k+2}^{-1} 
X_{T,2k+1}\\
 &=&BS_{2k+1} BS_{2k} &\qquad&&=&- S_{2k+2}S_{2k+1}\\
&&\mbox{for}\,k \in \{0\ldots p-1\}&&&& \mbox{for}\,k \in \{0\ldots p-2\},
\ea
$$
$$
P_{T-1,2p-1} = BS_{2p-1} \prod_{l = 1}^{2p-2} B_l \; BS_0
$$
and
$$
P_{T,2k}^2  = P_{T-1,2k+1}^2 = \unity.
$$
The difference of the vertex monodromies is
$$
P_{2T-1,2p-1} \ldots P_{T-1,1} P_{T,0}^{-1} \ldots  P_{T,2p-2}^{-1} = (-1)^{p-1} \unity.
$$
For defining the shift automorphism ${\cal S}_T$ we fix the 
roots $e^{i \xi_0(i P_{t,n}^B))} $
($t \in \{T,T-1\},n \in \{0 \ldots 2p-1\}$) as:
$$
e^{i \xi_0(i P_{T,2n}^B))} = e^{i \xi_k(i P_{T,2n}^B))} = 
\left( \frac{1+ k^2 (i P_{T,2n})^2 (-1)}
{k^2 + (i P_{t,n})^2 (-1)} \right)^{\frac{1}{2}} := \unity
$$
$ n \in \{0 \ldots p-1 \}$
and analogously
$$
e^{i \xi_k(i P_{T-1,2n+1}^B))} =  \unity
\qquad n \in \{0 \ldots p-1 \}
$$
\noindent f. a. $\quad k \in [0,1).$

\noindent Note that with this definition
\beqnao
P_{T-1,2p-1} \ldots P_{T-1,1} P_{T,0}^{-1} \ldots  P_{T,2p-2}^{-1} &=& (-1)^{p-1}
e^{-i \xi_0(i P_{T -1,2p-1}^B))} \ldots  e^{-i \xi_0(i P_{T-1,0}^B))} \\
&&e^{i \xi_0(i P_{T,0}^B))} \ldots e^{i \xi_0(i P_{T,2p-2}^B))}
\eeqnao
hence we have to be careful when defing the shift automorphism
(compare with (\ref{eq:MonX})).
\noindent Let
$$
c \dgleich \frac{2k}{1 + k^2}\qquad  b = -i \frac{1-k^2}{1+k^2},
$$
then using (\ref{theor:Rmatrix}) and normalizing, the local amplitudes
are straightforward obtained as: 
\beqnao
R_k(i P_{T,2n}) &=& \frac{i}{\sqrt{2 (1 - c)}} (b + (1-c) P_{T,2n})
\eeqnao
They obey the functional equations:
$$
\frac{R_k(i P_{T,2n})}{R_k(-i P_{T,2n})} = \frac{k+ i P_{T,2n}}{1+ k i P_{T,2n}};
$$
The evolution for the edge operators to the next time step is given by
$$
X_{T+1,n} = {\bf R}_T X_{T,n} {\bf R}_T^{-1} \quad \mbox{where} \quad 
{\bf R}_T = \prod_{l= 0}^{p-1} R_k(i P_{T,2l}).
$$
The shift matrix shall be given by:
$$
S_T^{-1} \dgleich R_0(i P_{T, 2p-2}) R_0(i P_{T-1,2p-3}) \ldots 
 R_0(i P_{T-1,1}) R_0(i P_{2T,0})
$$
which  defines the following shiftautomorphism on the edge algebra:
\beqnao
{\bf S}_T^{-1}(X_{T,n}) &:=& i  S_T^{-1} X_{T,n} S_T = X_{T,n-1} 
\qquad n \in \{1 \ldots 2p-1\} \\
{\bf S}_T^{-1}(X_{T,0}) &:=& i (-1)^{p-1} S_T^{-1} X_{T,0} S_T = X_{T,2p-1} 
\eeqnao
\begin{Lemma}
The matrices  $ {\bf C} P_{T,2n} {\bf C}^{-1} $ and 
 $ {\bf C} P_{T,2n+1} {\bf C}^{-1}$ $(n \in \{0 \ldots p-1\})$
commute with all generators
$I_{T}^{X}= $ $(X_{T,n})_{n \in \{0..2p\}}$ of ${\cal A}(X_T).$
\end{Lemma}
Hence 
\beqnao
&&{\bf R}_T  {\bf C} {\bf R}_T {\bf C}^{-1} {S}_T^{-1} 
{\bf C} { S}_T^{-1} {\bf C}^{-1} (X_{T,n} - i  {\bf C} X_{T,n}{\bf C}^{-1})
{\bf C} { S}_T  {\bf C}^{-1} { S}_T
{\bf C} {\bf R}_T^{-1}  {\bf C}^{-1} {\bf R}_T^{-1}\\
&=& 
{\bf R}_T {S}_T^{-1} X_{T,n} {S}_T {\bf R}_T^{-1}
- i {\bf C} {\bf R}_T {S}_T^{-1} X_{T,n} { S}_T {\bf R}_T^{-1}
{\bf C}^{-1}\\
&=& -i ({\bf R}_T  {\bf S}_T^{-1} (X_{T,n}) {\bf R}_T^{-1}
- i {\bf C} {\bf R}_T {\bf S}_T^{-1} (X_{T,n}) {\bf R}_T^{-1} {\bf C}^{-1})
\eeqnao
$n \in \{1 \ldots p-1\}$, analogous for $X_{T,0}$.

\noindent Clearly this defines lightcone shifts on the operators
$$
\psi_{T,n+1} \dgleich \frac{1}{2}(X_{T,n} - i 
{\bf C} X_{T,n} {\bf C}^{-1}) = \sigma^-_n \prod_{l = 0}^{n-1} \sigma_l^z,
$$
i.e.
$$
\psi_{T+1,n+1} \dgleich \tilde{{\bf R}}_{T} \tilde{{\bf S}}_{T}(X_{T,n})
 \tilde{{\bf R}}_{T}^{-1} 
$$
with $n \in \{0 \ldots p-1\}$
$$
\tilde{{\bf R}}_{T} \dgleich {\bf R}_T  {\bf C} {\bf R}_T {\bf C}^{-1}
\qquad
\tilde{{\bf S}}_{T}^{-1}(\psi_{T,n}) \dgleich i
{S}_T^{-1} {\bf C} { S}_T^{-1} {\bf C}^{-1} \psi_{T,n}
{S}_T {\bf C} { S}_T{\bf C}^{-1}
$$ 
and
$$
\sigma^{-} = \left( \ba{cc} 0&0\\1&0 \ea \right)
\qquad \sigma^z = B.
$$
The shift matrices $S_T$ and $S_T^{-1}$, as well as
the evolution matrix ${\bf R}_T$ are products of bilocal
operators 
$$
\ba{llll}
 R_k(i P_{T,2n})
&=& \unity \otimes \ldots R_k(i BS \otimes BS ) \ldots \otimes \unity
&n \in \{0 \ldots p-1\} \\
R_k(i P_{T-1,2n+1})
&=& \unity \otimes \ldots R_k(-i S \otimes S ) \ldots \otimes \unity 
& n \in \{0 \ldots p-2\}.
\ea
$$
Hence the same holds fot $\tilde{{\bf R}}_{T}$ and $\tilde{{ S}}_{T}$.
A straightforward computation gives

\beqnao
R_k(i BS \otimes BS ) {\bf C} R_k(i BS \otimes BS ) {\bf C}^{-1}
&=& R_k(i BS \otimes BS ) R_k(-i S \otimes S )\\
&=&  \unity \otimes \ldots 
\left( \ba{cccc} 1&0&0&0\\0&c&b&0\\0&b&c&0\\0&0&0&1 \ea \right)
 \ldots \otimes \unity =\\
=R_k(-i S \otimes S ) {\bf C} R_k(-i S \otimes S ){\bf C}^{-1}&&
\eeqnao
The shift matrices
$$
\tilde{S}_T^{-1} = \tilde{R}_0(iP_{T,2p-2}) \tilde{R}_0(iP_{T-1,2p-3})
\tilde{R}_0(iP_{T,2p-4}) \ldots \tilde{R}_0(iP_{T,0}) 
$$
act on the fermionic operators $\psi_{T,k}$ as
\beqna
&&\tilde{S}_T^{-1} \psi_{T,2n} \tilde{S}_T = \nonumber \\
&=&\tilde{R}_0(iP_{T,2n-2}) \psi_{T,2n}  \tilde{R}_0(iP_{T,2n-2})^{-1}\nonumber\\
&=& \left( \ba{cccc} 1&0&0&0\\0&0&-i&0\\0&-i&0&0\\0&0&0&1 \ea \right) 
(\sigma^-_{2n-1} \otimes \sigma^z_{2n-2}) 
\left( \ba{cccc} 1&0&0&0\\0&0&i&0\\0&i&0&0\\0&0&0&1 \ea \right) 
\prod_{l=0}^{2n-3} \sigma^z_l \nonumber\\
&=& -i \left( \ba{cccc} 1&0&0&0\\0&0&1&0\\0&1&0&0\\0&0&0&1 \ea \right) 
(\sigma^-_{2n-1} \otimes \sigma^z_{2n-2}) 
\left( \ba{cccc} 1&0&0&0\\0&0&1&0\\0&1&0&0\\0&0&0&1 \ea \right) 
\prod_{l=0}^{2n-3} \sigma^z_l \nonumber \\
&=& -i {\bf V}^{-1} \psi_{T,2n}  {\bf V}
\eeqna
$ n \in \{0 \ldots p-1\}$ with
$$
V_n = V_n^{-1} = \unity \otimes \ldots 
  \underbrace{\left( \ba{cccc} 1&0&0&0\\0&0&1&0\\0&1&0&0\\0&0&0&1 \ea \right)}_{(n+1,n)'th \; site}
 \ldots \otimes \unity \qquad
{\bf V}^{-1} =  V_{2p-2}^{-1} V_{2p-3}^{-1} \ldots V_{0}^{-1} 
$$
In a similar way 
\beqnao
\tilde{S}_T^{-1} \psi_{T,2n+1} \tilde{S}_T &=& -i {\bf V}^{-1} 
\psi_{T,2n+1}{\bf V}
\quad n \in \{1 \ldots p-1\} \\
\tilde{S}_T^{-1} \psi_{T,1} \tilde{S}_T &=& i (-1)^p {\bf V}^{-1} \psi_{T,1}{\bf V},
\eeqnao
so that finally
$$
\ba{lllll}
\tilde{{\bf S}}_T^{-1}(\psi_{T,n}) &=& 
 i \tilde{S}_T^{-1}\psi_{T,n}\tilde{S}_T &=& {\bf V}^{-1} \psi_{T,n} {\bf V}
\qquad n \in \{2 \ldots 2p\}\\ 
\tilde{{\bf S}}_T^{-1}(\psi_{T,1}) &=& 
 i (-1)^{p-1}\tilde{S}_T^{-1}\psi_{T,n}\tilde{S}_T &=& {\bf V}^{-1} \psi_{T,1} {\bf V}\ea
$$
which is identical to the shift automorphism constructed in \cite{DV}.
Following (\ref{prop:Shif},\ref{prop:Shif2}) we know
that the construction of the shift automorphism $\tilde{S}$ and the
evolution automorphism given by the conjugation with $\tilde{R}$
is sufficient for constructing a hamiltonian quantum evolution in
the sense of the previous sections.

Finally  comparing with the construction in \cite{DV} one finds 
 that the fermionic operators obey an evolution of
free massive fermions. The evolution equations can be derived easily
by considering the evolution for the edge variables.
Remembering (\ref{eq:idem}) one finds:
\beqna
X_{t+1,2k} \dgleich {\bf R}_t X_{t,2k} {\bf R}_t^{-1}
&=& \frac{k+ i X_{t,2k+1}^{-1} X_{t,2k}}{1- k i X_{t,2k+1}^{-1} X_{t,2k}}
\frac{1- k i X_{t,2k+1}^{-1} X_{t,2k}}{1- k i X_{t,2k+1}^{-1} X_{t,2k}}
X_{t,2k} \nonumber \\
 &=& \frac{1}{1+k^2}(2k + i(1-k^2)  X_{t,2k+1}^{-1} X_{t,2k}) X_{t,2k} 
\nonumber \\
&=& c X_{t,2k} + b X_{t,2k+1}
\eeqna
analogously for $X_{t+1,2k+1}$.
Since these equations are linear,
the evolution equations for the fermionic operators follow immediately:
\beqna
\psi_{t+1,2n-1} &=& c \psi_{t,2n-1} + b \psi_{t,2n}\\
\psi_{t+1,2n} &=& c \psi_{t,2n} + b \psi_{t,2n-1}.
\eeqna

\subsection{Conclusion}

In the present paper a generalized model of a lattice field theory
of sine Gordon type at root of unity was suggested. Among others
the aim was to stress the purely local character of the evolution
automorphism and finally to derive global features like constant 
transfermatrices and hence a connection to models in statistical mechanics
by restricting to a special case
 ( please refer \ref{prop:Shif},\ref{prop:Shif2}) or by
considering symmetries of the models in consideration. 

Since the classical phase space belonging to the evolution of e.g. the
face variables (current variables) is a Torus $S^{2p}$ the corresponding
quantum model can be viewed as a kind of quantization of this 
torus. Hence a next future project should be the investigation
of the above within the framework of noncommutative geometry
\cite{Co}. 

A nice side effect of the study of the above quantum lattice model
was the detection of
a relation to another quantum lattice model, namely the
massive Thirring model in it's reduced version as describing 
free massive fermions, as given by \cite{DV}. Since relations
between these two models are known for the continous case, see
e.g. \cite{KM,C} it seems to speak for the self coherence of 
the above lattice models, that they also exist in the
discrete case.

\section{Acknowledgements}

My heartily thanks go to  Ulrich Pinkall, Ruedi Seiler and
Michael Karowski for very helpful
discussions and for pointing out reference \cite{DV} and \cite{TS}.

{}

\end{document}